\documentclass{aa}
\usepackage{txfonts}
\usepackage{natbib}
\usepackage{graphicx}
\usepackage{longtable}
\bibpunct{(}{)}{;}{a}{}{,}

\begin{document}

\title{Multiwavelength campaign on Mrk~509 VII. Relative abundances of the warm absorber}

\author{K. C.~Steenbrugge\inst{1,2}
 \and J. S. Kaastra\inst{3,4}
 \and R. G. Detmers\inst{3,4}
 \and J. Ebrero\inst{3}
 \and G. Ponti\inst{5}
 \and E.~Costantini\inst{3}
 \and G. A. Kriss\inst{6,7}
 \and M. Mehdipour\inst{8}
 \and C. Pinto\inst{3}
 \and G. Branduardi-Raymont\inst{8}
 \and E. Behar\inst{9}
 \and N. Arav\inst{10}
 \and M. Cappi\inst{11}
 \and S. Bianchi\inst{12}
 \and P.-O. Petrucci\inst{13}
 \and E. M. Ratti\inst{3}
 \and T. Holczer\inst{9}
}

\offprints{katrien.steenbrugge@gmail.com}

\institute{ 
	 Instituto de Astronom\'{i}a, Universidad Cat\'{o}lica del Norte, Avenida Angamos 0610, Antofagasta, Chile 
	 \and
         Department of Physics, University of Oxford, Keble Road, Oxford OX1 3RH, UK
         \and
	     SRON Netherlands Institute for Space Research,
Sorbonnelaan 2,
                3584 CA Utrecht, the Netherlands
         \and
                Astronomical Institute, Utrecht University, P.O. Box
80000,
                3508 TA Utrecht, the Netherlands
	         \and 
             School of Physics and Astronomy, University of Southampton, Highfield, Southampton SO17 1BJ, UK
         \and 
             Space Telescope Science Institute, 3700 San Martin Drive, Baltimore, MD 21218, USA
         \and 
             Department of Physics \& Astronomy, John Hopkins University, Baltimore, MD 21218, USA
         \and
             Mullard Space Science Laboratory, University College London, Holmbury St Mary, Dorking, Surrey, RH5 6NT, UK
         \and
             Department of Physics, Technion, Haifa 32000, Israel
         \and
             Department of Physics, Virginia Tech, Blacksburg, VA 24061, USA
         \and
             INAF-IASF Bologna, Via Gobetti 101, I-40129 Bologna, Italy
         \and
             Dipartimento di Fisica, Universit\'{a} degli Studi Roma Tre, via della Vasca Navale 84, I-00146 Roma, Italy
         \and
             UJF-Grenoble 1 / CNRS-INSU, Institut de Plan\'{e}tologie et d’Astrophysique de Grenoble (IPAG) UMR 5274, Grenoble, F-38041, France}

\date{\today}

\abstract
{The study of abundances in the nucleus of active galaxies allows us to investigate the evolution of abundance by comparing local and higher redshift galaxies. However, the methods used so far have substantial drawbacks or rather large uncertainties. Some of the measurements are at odds with  the initial mass function derived from the older stellar population of local elliptical galaxies. }
{We determine accurate and reliable abundances of C, N, Ne, and Fe relative to O from the narrow absorption lines observed in the X-ray spectra of Mrk~509.} 
{We use the stacked 600~ks XMM-{\it Newton} RGS and 180~ks {\it Chandra} {\rm LETGS spectra. Thanks to simultaneous observations with INTEGRAL and the optical monitor on-board XMM-}{\it Newton} {\rm for the RGS observations and HST-COS and Swift for the LETGS observations, we have an individual spectral energy distribution for each dataset. Owing to the excellent quality of the RGS spectrum, the ionisation structure of the absorbing gas is well constrained, allowing for a reliable abundance determination using ions over the whole observed range of ionisation parameters.} }
{We find that the relative abundances are consistent with the proto-solar abundance ratios: C/O = 1.19$\pm$0.08, N/O = 0.98$\pm$0.08, Ne/O = 1.11$\pm$0.10, Mg/O = 0.68$\pm$0.16, Si/O = 1.3$\pm$0.6, Ca/O = 0.89$\pm$0.25, and Fe/O = 0.85$\pm$0.06, with the exception of S, which is slightly under-abundant, S/O = 0.57$\pm$0.14. Our results, and their implications, are discussed and compared to the results obtained using other techniques to derive abundances in galaxies.}
{}



\titlerunning{Abundances of the warm absorber in Mrk~509}
\authorrunning{K. C. Steenbrugge et al.}

\maketitle

\section{Introduction}
The X-ray emission from active galactic nuclei (AGN) is generally thought to be Comptonized emission originating from near the accretion disk surrounding the supermassive black hole, which accretes gas from the host galaxy. In general, the continuum is well described by a power-law component and a soft excess. In about 50\% of Seyfert 1 galaxies \citep{crenshaw03}, narrow absorption lines from highly ionised gas are observed, with the gas generally outflowing at a velocity of between 100 and a few 1000 km s$^{-1}$ \citep{kaastra02}. This absorption can be studied in detail in the UV and X-ray part of the spectrum. The spectral resolution is significantly higher in the UV, but only a few absorption transitions are found in this part of the spectrum. Therefore, determining the ionisation structure of the plasma is difficult. In contrast, in the X-ray band a multitude of ions have observable transitions, spanning about four orders of magnitude in ionisation parameter \citep{steenbrugge05, holczer07}, but the spectral resolution is significantly poorer and the different velocity components cannot be fully distinguished. The most reliable results are therefore obtained by performing a joint analysis of the UV and X-ray spectra, preferentially obtained simultaneously \citep{steenbrugge05,costantini07,costantini10}, so that the velocity structure obtained from the UV spectrum can be combined with the ionisation structure determined from the X-ray spectrum. 

Although several detailed studies of warm absorbers (i.e. the ionised outflow observed through narrow absorption lines in the X-ray spectrum) have been accomplished in the past few years, the geometry and distance of the gas components constituting the warm absorber are only now being determined. We therefore executed a large observing campaign targeting the Seyfert 1 galaxy Mrk~509 to help us resolve these questions \citep{kaastra11a}. The absorber is generally assumed to originate either from the accretion disk or the obscuring torus, thus the abundances determined should closely match those of either the accretion disk or the obscuring torus. The abundances determined should be representative of the abundances in the vicinity of the supermassive black hole and the nucleus of Mrk~509. 

In the case of Mrk~509, the soft excess originates from an optically thick plasma with a temperature of 0.2~keV \citep{mehdipour11}, while the hard X-ray power-law likely originates in an optically thin plasma with a much higher temperature \citep{paltani11}. An absorber is detected with a velocity range between $-$426 and +219 km~$^{-1}$ \citep{kriss11}. 

Determining the relative abundances in the nucleus allows us to study the enrichment processes in the host galaxy of Mrk~509. Furthermore, the relative abundances that we measure can be compared with the (relative) abundances measured for active galactic nuclei (AGN) at high redshifts, to determine the abundance evolution of AGN and thus the history of the enrichment processes prevalent at different epochs. At high redshifts, abundances are usually determined from the relative strength of two or more broad emission lines shifted into the optical band. This broad-emission-line gas is  emitted within 0.1~pc of the central black hole \citep{peterson00}. In Mrk~509, the size of the broad line region (BLR) determined from H$\beta$ measurements is 80~light-days (0.067~pc) \citep{carone96}. The absorbing gas is probably located at a somewhat larger distance, although both aspects of the AGN phenomenon are likely to have similar abundances because of their close origin. Using this assumption allows one to compare the abundances determined by these different methods, and thereby the abundances derived for different redshifts. 

The main advantage of using X-ray absorption lines over the optically detected broad emission lines is that the ionisation structure of the absorber can be accurately determined, which is crucial in differentiating the effects on the plasma of the abundances from those of the ionisation structure. Unless the ionisation structure is well known, one can only consider ions that have their peak ionic column density at the same ionisation parameter to determine the abundances. There are no hydrogen transition lines in the X-ray spectra; only continuum absorption is present. However, the bound-free absorption from H is degenerate with the continuum model used, which is not a priori known. Therefore, only relative abundances are determined throughout this paper. In addition, unless mentioned otherwise, all the relative abundances are with respect to proto-solar ones (\citealt{lodders09}, see the last column in our Table~\ref{tab:abun1}). These abundances were determined from chondritic meteorites and the photospheric abundances of the Sun. Here we measure the abundances of C, N, Ne, Mg, Si, S, Ca, and Fe relative to O for gas analysed based on its narrow absorption lines in the stacked XMM-{\it Newton} reflection grating spectrometer (RGS) and {\it Chandra} low energy transmission grating (LETGS) X-ray spectra of Mrk~509. We thus determine the relative abundances of those elements that result from the different enrichment processes occurring in the host galaxy. 

In Sect.~2, we give a brief overview of the data and spectral models used, while in Sect.~3 we present the results. In Sect.~4, we discuss the obtained relative abundances, and compare these to the abundances quoted in the literature and derived using different methods. Our conclusions are given in Sect.~5.

\section{Observations and spectral models}
\subsection{Observations}
The data used in this study were obtained to determine the distance from the ionising source of the absorbers in Mrk~509 by measuring the time delay between the change in flux and ionisation structure of the gas. This goal constrained the observing strategy, which is described in detail by \cite{kaastra11a}. We present here a brief summary of the X-ray data used in this article, which consists of ten $\sim$60~ks XMM-{\it Newton} observations taken four days apart on October$-$November 2009. The 180~ks LETGS ({\it Chandra}) spectrum was obtained 20 days later over two consecutive orbits between Dec. 10 and 13, 2009. 

Full details of the RGS data reduction, calibration, and stacking of the data are given in \cite{kaastra11b} and for the LETGS data by \cite{ebrero11}. We note that the spectra were aligned using the very accurate aspect solution of the satellite before the stacking occurred, so that no line from the source is smeared out, and that special care was taken not to introduce weak emission or absorption features caused by a bad pixel in one of the spectra.

\subsection{Spectral models}

Here we summarise the absorption models given in \cite{detmers11} and \cite{ebrero11} which we used to determine the relative abundances. The continuum was fit by a logarithmically spaced spline function, the advantage being that this model does not make any a priori assumptions about the physical components of the continuum. Several narrow and broad emission lines were fit, as well as a few radiative recombination continua (RRC). Together these components allow for an accurate continuum fit, necessary to perform accurate line depth measurements. The local Galactic foreground absorption is fit with three collisionally ionised {\tt hot} components with different temperatures. The {\tt hot} model gives the transmission of a layer of collisionally ionised gas. Free parameters are the hydrogen column density, the abundances, and the temperature. 

The AGN absorption observed in the stacked RGS spectrum was fit with several different models. Model 1 includes two {\tt slab} components, while model 2 uses three {\tt slab} components. The {\tt slab} model gives the transmission of a layer with arbitrary ionic composition. The main free parameters are the ionic column densities, but also the velocity outflow and velocity broadening are free parameters. In model 2 the ionic column density is determined for all ions for the three velocity components. The ionic column densities obtained with model 1 were used as the input to the absorption measure distribution (AMD) modelling, which is the result we use in this paper. Model 3 fits six xabs components. The {\tt xabs} model gives the transmission of a photoionised layer, where the ionic column densities are determined from the total hydrogen column density, the abundances, and the ionisation parameter.  For the LETGS, similar models for the absorber were used, although, for the {\tt xabs} and AMD modelling only three components were detected significantly in the spectrum, owing to the lower signal-to-noise ratio (hereafter SNR). Owing to the complexity of the models used, one of which, model 2 for the RGS, has 22 components and nearly 100 free parameters, we decided not to fit the RGS and LETGS spectrum simultaneously. We instead fitted the abundances for the RGS and LETGS data separately, and compare in Sect.\ref{sect:comprl} the values obtained for the separate analyses. 

To obtain an acceptable fit to the RGS spectrum, we need at least three velocity components, and find consistent results for the LETGS spectrum. We note that for Mrk~509, seven different velocity components are detected in the FUSE spectrum \citep{kriss00}, eight in the HST-STIS UV spectrum \citep{kraemer03}, and thirteen in the HST-COS spectrum \citep{kriss11}, with velocities ranging between $-$426 and +219~km s$^{-1}$. The current X-ray spectra do not have the spectral resolution to separate these different velocity components. However, in the X-ray band we only detect absorption from velocities that are consistent with being at rest with the host galaxy or outflowing. The absorption component with the highest redshift relative to the systemic velocity is only observed in the UV and is likely to be a high velocity cloud near the host galaxy \citep{kriss11}. From studying the absorption troughs either at rest or outflowing in the UV, it appears that these are clustered around $-$300 and +16~km s$^{-1}$ \citep{kriss00,kraemer03,kriss11}. Fitting model 2, \cite{detmers11} finds outflow velocities of $-$13$\pm$11~km~s$^{-1}$ and $-$319$\pm$14~km~s$^{-1}$, consistent with the main absorption troughs detected in the UV. \cite{detmers11} also detects a higher velocity component at $-$770$\pm$109~km~s$^{-1}$, but observed in only two ions: \ion{Mg}{xi} and \ion{Fe}{xxi}, that is not present in the UV spectra. Fitting the six different {\tt xabs} models for the RGS, \cite{detmers11} indeed find that there is a slow (fit by components B1 and C1) and a faster (fit by components A2, C2, D2 and E2) component, with different outflow velocities. In their model 1, \cite{ebrero11} used two velocity components in the {\tt slab} modelling similar to model 1 for the RGS. Owing to a difference in absolute wavelength calibration between the RGS and the LETGS the two velocities are: $-$338~km s$^{-1}$ and +3~kms$^{-1}$, which are within the uncertainty consistent with the velocities fitted for the RGS and the two UV detected troughs.

The stacked 600~ks high SNR RGS spectrum was analysed by \cite{detmers11}. In this paper, we use the results from their model 3, which has six {\tt xabs} components and is detailed in their Table~5, and AMD fit (see Table~6 in \citealt{detmers11}), to determine the relative abundances. The latter is based on the results of model 1. For the LETGS, we use the results of the {\tt xabs} modelling obtained by \cite{ebrero11}, as well as the AMD modelling. The only parameters we allowed to vary in our fits were the abundances of C, N, O (in the AMD fit), Ne, Mg, Si (only for the LETGS spectrum), S, Ca, and Fe, and the hydrogen column density. We thus kept the ionisation parameter, or the ionic column densities for the AMD fit, as well as the outflow velocity and the velocity broadening fixed to the best-fit values determined by \cite{detmers11} and \cite{ebrero11}. We did test whether leaving the ionisation and velocity parameters free significantly alters the derived relative abundance of S, but found that these were unchanged to one standard deviation, namely a change from 0.57 (see Table~\ref{tab:abun1}) to 0.63, where the determined error in S abundance is 0.14 for the {\tt xabs} modelling of the RGS. For the AMD modelling, we used the fitted ionic column densities as input parameters, which hence have to be kept fixed. However, these ionic column densities were determined without assuming either an ionisation structure or abundances for the plasma. We briefly note that the continuum was significantly lower during the LETGS observation, 24.19 versus 37.78 in 10$^{-14}$ W~m between 5~\AA~and 40~\AA~\citep{kaastra11a}. In addition to the shorter exposure time, this leads to a lower SNR in the LETGS spectrum. Therefore, the best-fit absorption and emission model is necessarily less detailed than that obtained for the RGS spectrum. All the spectral analysis is done with the {\tt SPEX} software\footnote{http://www.sron.nl/spex} \citep{kaastra06}, and the errors quoted are 1$\sigma$. 

\section{Derived abundances}
Hydrogen only produces continuum absorption, which is degenerate with our continuum parameters. Thus determining the absolute abundances, i.e. the abundances relative to hydrogen, which would likely be in the form of \ion{H}{ii}, is difficult. Hence, we used abundance ratios where we chose an element with a large number of absorption lines over a large ionisation range and determined the abundances relative to this element. For this element, we used oxygen because it is the most abundant metal and we observe transitions from \ion{O}{iv} to \ion{O}{viii} with multiple detected lines per ion, with the exception of \ion{O}{iv}. 

\subsection{RGS determined abundances \label{sect:rgs}}
\subsubsection{Abundances determined assuming all components have the same abundances}
The stacked RGS spectrum has a much higher SNR than LETGS and thus provides us with more accurate relative abundances. We use model 3, which fits the absorption with six {\tt xabs} components that each have a different outflow velocity and the AMD modelling to determine the relative abundances. For the parameters used in these models we refer to \cite{detmers11}, who assumed the proto-solar abundances given by \cite{lodders09}. Here we determine the relative abundances using three methods based on model 3, the AMD model, and different assumptions. 

We first use model 3 and assume that all the differently ionised and different kinematic components have the same abundances. We allow the abundances of C, N, Ne, Mg, S, Ca, and Fe to vary and choose O as the reference element. Since no Si absorption line is visually detected in the RGS spectrum, for the RGS abundance determination we did not fit the Si abundance. We thus couple the abundances for the six {\tt xabs} components. The resulting relative abundances are given in column 2 of Table~\ref{tab:abun1}. From Table~\ref{tab:abun1}, we conclude that the abundance ratios are consistent with the proto-solar ratios with the exception of the S/O. 

\begin{table*}
\begin{center}
\caption{Best fit abundances relative to oxygen compared to the proto-solar abundance ratio from the RGS spectrum (using model 3 and the AMD models, see Sect~\ref{sect:rgs}) and the LETGS spectrum using the {\tt xabs} and discrete AMD model (see Sect~\ref{sect:letgs}). We assume that the abundances for the faster and slower outflow component are the same. }
\label{tab:abun1}
\begin{tabular}{llll|ll|c|ll|ll}
\hline
\hline
       & RGS:           &               &               & LETGS:       &                    &              &       &              &  & \\     
ion    &  model 3    & discrete AMD  & continuous AMD& {\tt xabs}   & discrete AMD$^{a}$ & recommended$^{b}$& ion$^{c}$&              & ion$^{d}$    &   \\\hline
C/O    & 1.19$\pm$0.08  & 1.00$\pm$0.07 & 1.01$\pm$0.03 & 1.2$\pm$0.3  & 1.00               & 1.19$\pm$0.08 & C/Fe & 1.40$\pm$0.1 & C            & 8.46   \\
N/O    & 0.98$\pm$0.08  & 0.80$\pm$0.05 & 0.80$\pm$0.03 & 0.5$\pm$0.2  & 0.80               & 0.98$\pm$0.08 & N/Fe & 1.15$\pm$0.06& N            & 7.90    \\
O      &                & 1.00$\pm$0.05 & 1.00$\pm$0.05$^{e}$ &        & 1.11               &               & O/Fe & 1.17$\pm$0.06& O            & 8.76  \\
Ne/O   & 1.11$\pm$0.10  & 1.17$\pm$0.15 & 1.16$\pm$0.07 & 0.9$\pm$0.3  & 1.17               & 1.11$\pm$0.10 & Ne/Fe& 1.31$\pm$0.17& Ne           & 7.95   \\
Mg/O   & 0.68$\pm$0.16  & 0.92$\pm$0.29 & 0.93$\pm$0.15 & 0.7$\pm$0.6  & 2.9$\pm$5.3        & 0.68$\pm$0.16 & Mg/Fe& 0.80$\pm$0.16& Mg           & 7.62  \\
Si/O   &                &               &               & 1.3$\pm$0.6  & 1.2$\pm$1.5        & 1.3$\pm$0.6   & Si/Fe& 1.53$\pm$0.60& Si           & 7.61   \\
S/O    & 0.57$\pm$0.14  & 0.68$\pm$0.24 & 0.69$\pm$0.09 & 0.4$\pm$0.4  & 1$\pm$1.1          &  0.57$\pm$0.14 & S/Fe &  0.67$\pm$0.15 & S            & 7.26  \\
Ca/O   & 0.89$\pm$0.25  & 3.34$\pm$0.75 & 3.81$\pm$0.86 & 2.7$\pm$1.6  & 5.1$\pm$4.7        & 0.89$\pm$0.25 & Ca/Fe& 1.05$\pm$0.26& Ca           & 6.41  \\ 
Fe/O   & 0.85$\pm$0.06  & 0.81$\pm$0.05 & 0.79$\pm$0.08 & 1.1$\pm$0.2  & 0.81               & 0.85$\pm$0.06 & Fe   & 
        & Fe           & 7.54 \\\hline
\end{tabular}\\
$^{a}$ Values without error bars denote that they were kept fixed to the best fit discrete AMD model of the RGS spectrum.\\
$^{b}$ The most accurately determined relative abundances that we use in the discussion of this paper.\\
$^{c}$ The calculated relative abundances relative to Fe, for comparison with other measurements.\\
$^{d}$ Logarithmic proto-solar abundances of \cite{lodders09}, where H has a value of 12.\\ 
$^{e}$ Fixed to the value obtained for the discrete AMD model.
\end{center}
\end{table*} 


For iron, which has a best-fit value 2.5$\sigma$ below the proto-solar ratio, some of the stronger lines are notably weaker in the best-fit model, but the main difference is in the absorption due to the unresolved transition array (UTA between 16 and 17.5~\AA, \citealt{sako01,behar01}), because the many absorption lines result in pseudo-continuum absorption. The spacing of 2.68~\AA~and 3.18~\AA~between the points of the spline used to fit the continuum in the range  14$-$18~\AA~is wider than the UTA, which has a width of about 1.5~\AA. 

\subsubsection{Abundances of each velocity component}

Thus far we have assumed that the faster and slower velocity components and the different ionisation components have the same abundance. The latter assumption seems justified, if the different ionisation components are part of the same outflow and thus form one physical entity with a common origin. The former assumption is more questionable if the different velocity components are from different outflow structures formed at different distances and possibly with a different origin. The assumption that all outflow components have the same abundance was also made by \cite{arav07} in their study of the abundances of Mrk~279 using FUSE and HST-STIS UV spectra. The lower velocity outflow component might be contaminated by absorption from the host galaxy. However, as we observe the host galaxy face on and detect no absorption from neutral gas in the X-ray spectra at the redshift of Mrk~509, this should have a limited effect on the derived abundances. However, we note that in the UV substantial \ion{C}{iii} and \ion{Si}{iv} is detected from the host galaxy \citep{kriss11}, and there could thus be contamination, certainly for the more lowly ionised gas. Another possible contaminant is the inflowing gas seen in UV absorption lines \citep{kriss11}. We note that in the UV detected \ion{O}{vi} line the strongest component has a velocity of +219~km~s$^{-1}$ but this velocity component is not detected in the RGS spectrum. Therefore, the contamination by absorption from inflowing gas is also likely to be limited. 

Considering the small error bars in the determined abundances, we decided to attempt, as a second method the determination of the abundances separately for the fast and slow outflow components. We still tied the abundances for the different ionisation components for each velocity component. We again use O as a reference element. We thus fit two abundances per element simultaneously. As expected, the error bars are larger owing to the correlations between both velocity components. 


With the exception of the Fe/O ratio for the slow velocity component, all abundance ratios are consistent with proto-solar ratios and the values determined assuming the abundance is the same for both velocity components. The only difference is for iron, but this might be due to the difficulty in fitting the many weak and lowly ionised lines, as well as some uncertainty in the centroids of some of the blends. 


\subsubsection{Abundances from the absorption-measure distribution models}

Since the effective spectral resolution is insufficient to separate the slow and faster velocity components into independent components, and that likely part of the fastest outflow velocity detected in just two ions is improperly fit with the six {\tt xabs} components, we decided to use the ionic column densities determined from model 1 \citep{detmers11} to derive the absorption measure distribution and the abundances simultaneously. The procedure used to determine the AMD taking into account the full ionisation range at which an ion is formed is described by \cite{detmers11}. 

We used in this method, in contrast to the previous methods, the ionic column densities measured from the spectrum using two {\tt slab} components. Thus, a priori we did not assume any ionisation structure, which in model 3 is assumed to be discrete and determined by the {\tt xabs} components. To obtain the relative abundances from the measured ionic column densities, we did need to assume a certain SED and ionisation balance, which were taken to be the same as for the {\tt xabs} modelling. An advantage of this method is that we can test whether the ionisation structure is discrete or continuous, as suggested for NGC~5548 by \cite{steenbrugge05}. We thus determined the abundances for both discrete and continuous ionisation models. The details of these fits are described in \cite{detmers11}. Since no hydrogen lines are present, we were unable to determine absolute abundances. We determined the abundances of individual elements relative to all other elements with measured column densities, because the fit first derives the total column density assuming proto-solar abundance ratios and the fit is then refined by adjusting the abundances. As before, we therefore re-normalise our abundances to oxygen. The error bar given for oxygen is the nominal uncertainty in the oxygen abundance relative to all other elements.

The abundances for a discrete and a continuous AMD model are given in columns 3 and 4, respectively, of Table~\ref{tab:abun1}. Consistent abundances were obtained by assuming either a discrete or a continuous ionisation structure. This provided us with confidence that we can determine the ionisation structure of the gas accurately enough to prevent any influence on our abundance determination. \cite{detmers11} does convincingly show that the ionisation structure is discrete, at least for highly ionised gas, hence we only discuss this model further here.

\subsection{LETGS determined abundances \label{sect:letgs}}
Here we repeated two of the above mentioned methods to measure the abundance, but this time for the {\it Chandra} LETGS data. Owing to the lower SNR of the data, we used C-statistics to fit the data, instead of $\chi^2$ statistics as was the case for the RGS data. We determined the abundances using the spectral model with three {\tt xabs} components described by \cite{ebrero11} assuming the continuum is well described by a spline. The difference in absorption parameters are negligible if a power-law and modified black body were assumed instead of a spline. The outflow velocity and the velocity width were fitted separately for each {\tt xabs} component, as done for the RGS analysis. Owing to the lower SNR, only three instead of six {\tt xabs} components are significantly detected \citep{ebrero11}. This is in part because the slower and faster outflow components are only partially resolved and because the lowest ionisation component is not significantly detected. The lowest ionisation component also has the lowest hydrogen column density, and is thus more difficult to detect significantly in lower SNR data. We do know that low ionisation gas was present during the LETGS observation, because it was detected in the HST-COS spectra \citep{kriss11}. The best-fit abundances that we derived using three {\tt xabs} components, and tying the abundances for the different {\tt xabs} components, for the LETGS spectrum are listed in column 5 of Table~\ref{tab:abun1}. 


We also determined the abundances using the AMD modelling, assuming a discrete ionisation structure, using the ionic column densities determined for model 1, which includes two {\tt slab} components \citep{ebrero11}. This model is the same as model 1 for the RGS and has two different velocity components, although each ion is only fit using one velocity component. The AMD fit only detects three discrete ionisation components, owing once again to the lower SNR of the data set, and consistent with the {\tt xabs} modelling. Considering the lower SNR, we assumed that the ionisation structure of the gas is discrete, as this gave a more accurate description of the RGS data than the continuous model. However, the fit is unstable if we allow the ionisation parameters, hydrogen column densities, and abundances to be free parameters, owing to the strong correlations between these parameters caused by the limited SNR. The error bars in the ionic column densities are too large, thus we decided to fix the abundances for those elements that have a well-determined abundance from the AMD analysis of the RGS data (C, N, O, Ne, and Fe). However, owing to the larger wavelength range, we can determine the Si abundance, and possibly more tightly constrain the Mg, S, and Ca abundances.  



The uncertainties in both methods are generally quite large for the ions for which we allowed the abundance to vary, because these elements show only a few weaker lines and the abundances were also poorly determined with the RGS spectrum. The abundances of these elements (Mg, Si, S, and Ca) are consistent with proto-solar abundances. The improvement in the fit with the best-fit abundances and the proto-solar abundances using xabs modelling is small, indicating that they are not very secure determinations, as can also be seen from the error bars and the difference in the values derived by the {\tt xabs} and AMD modelling. As a result of the larger error bars, the relative abundances are consistent with those determined for the RGS spectrum. 

Since the higher SNR and the spectral resolution of the RGS spectrum are too poor to accurately determine the abundances for each velocity component, we did not try to fit the abundances for each velocity component in the LETGS spectrum.  

\section{Discussion}

We used the narrow absorption lines to measure the relative abundances, but instead of using the UV-detected narrow absorption lines, as \cite{arav07} did, we used the X-ray detected narrow absorption lines. An advantage of using absorption lines is that the lines, with the exception of very weak and in the case of Mrk~509 undetected metastable lines, are independent of the density and temperature of the gas. 

The use of X-ray, instead of UV, detected absorption lines has several advantages. One advantage is that there are absorption lines from enough ions to accurately determine the ionisation structure of the absorbing gas. The many iron and oxygen ions detected in X-ray spectra enable the ionisation structure to be accurately determined using only the oxygen or iron ions. This can be done without abundance differences complicating the analysis, something that is impossible from the lines detected in the UV spectrum. The analysis of UV spectra is also complicated by the saturation of the absorption lines and the fact that one has to correct for non-black saturation and a velocity-dependent covering factor \citep{arav03}. In the X-ray spectrum of Mrk~509, and for most of the other Seyfert 1 galaxies, the covering factor is unity \citep{detmers11,steenbrugge09}. We did test whether the covering factor is indeed unity by forcing the covering factor of the six {\tt xabs} components to be 0.7 and refitting the RGS spectrum, allowing the hydrogen column densities and ionisation parameters of the different {\tt xabs} components to vary as free parameters. The resultant $\chi$$^2$ increased by 55 from the best fit with unity covering factor. If we then let also the covering factors free, for all six components this resulted in a best fit with unity covering factors. Furthermore, in X-ray spectra there are enough line series and weaker lines to correctly determine the column density of the ions, even if the deepest lines are saturated. We should note that in Mrk~509 the lack of detection of the weaker absorption lines observed in for instance NGC 5548 \citep{steenbrugge05} indicates that generally, i.e. with the exception of the strongest lines, the absorption lines detected are not saturated. A final advantage is the large range in ionisation parameter sampled in the X-ray band, allowing us to study many ions of an element when determining its abundance. 

However, there are also disadvantages. One is that we cannot separate the velocity components detected in UV spectra, thus as in abundance determinations based on UV spectra, we have to assume that all velocity components have the same abundances. Another disadvantage is that in the X-ray band no transitions of H are observable, thus the determined abundances are all relative. Only FUSE enabled the full Lyman series to be observed for this redshift range, permitting absolute abundances of AGN to be determined.

The measured abundances relative to oxygen are consistent with proto-solar abundance ratios with the exception of sulphur. For S, the different methods and the different instruments lead to a result implying that S is more likely to be under-abundant than have a proto-solar abundance ratio. For this element, we did test whether allowing the ionisation and velocity parameters to vary in the fit that determines the abundances has a significant effect on the determined S abundance ratio. However, the additional free parameters increase the abundance by less than 1~$\sigma$ for the RGS spectrum fit with the six {\tt xabs} components. The abundance is determined from three lines longward of 31~\AA, and it is thus a fairly robust result. Sulphur is formed mostly in supernovae (hereafter SNe) type Ia, as are Ca and Fe, although SNe~II are a far more significant contributor to the S abundance than to either the Ca and Fe abundance. If we study the abundance ratios relative to iron rather than oxygen, then S has an abundance ratio consistent with proto-solar, but carbon is slightly overabundant.

\subsection{Comparison between the abundances derived \label{sect:comprl} }

All the abundance ratios, apart from that of Ca, derived for the RGS spectrum are consistent between the different methods used. The differences between the relative abundances for the continuous versus discrete AMD and the abundances derived using model 3, provide a good indication of the systematic errors caused by the different ionisation structure modelling. Since the abundance ratios inferred from the AMD modelling are consistent with those derived using model 3, owing to the error bars being on average larger, for the remainder of the paper we use the abundances determined from model 3, assuming that all the velocity components have the same abundances.

The abundance measurements are consistent to within 2.5$\sigma$, and there is no general trend observed for the results of the RGS and LETGS data sets. The ions fit for both the RGS and LETGS data using the AMD model give abundances that are consistent to within 1$\sigma$. This is unsurprising, as we only fit those ions that had larger error bars in the RGS abundance determination, in the LETGS spectral fitting. 

The Ca/O abundance measurements are slightly inconsistent to within 3$\sigma$ between the different methods used to determine the abundance from the RGS spectrum. The {\tt xabs} measured abundance is consistent with the proto-solar Ca/O ratio, while the measured AMD abundance is inconsistent with the proto-solar ratio. In both the RGS and LETGS spectra, only a \ion{Ca}{xiv} absorption line is visually detected. This is also the only ion that is detected in the {\tt slab} modelling (models 1 and 2) of the absorber. The detected \ion{Ca}{xiv} line is blended with absorption from the local \ion{N}{vii}~Ly$\alpha$ line. Owing to this blending and since no \ion{Ca}{xiii} or \ion{Ca}{xv} lines are detected, and the {\tt xabs} modelling is more reliable in determining the column densities of weak and blended features, we use the RGS data {\tt xabs} abundance in the rest of this paper.

\subsection{Comparison between our abundances and nuclear abundances in the literature}
The chemical abundances of quasars provides insight into the evolution of their host galaxy and in particular the enrichment processes such as star formation. Therefore, the abundances of high redshift quasars especially have been well-studied. The two principal methods for measuring quasar abundances are the study of the broad emission lines \citep{shields76,hamann99} and the study of the narrow absorption lines \citep{gabel06,arav07}. The broad emission lines are formed in the vicinity of the supermassive black hole at distances from the black hole generally between 1$-$70 light-days (8.4$\times$10$^{-4}$ $-$ 0.059~pc) \citep{peterson00}. 

The outflow also originates close to the super-massive black hole, as it is released either by the accretion disk or the torus.
Hence, both methods determine the abundances in the vicinity of the central black hole. Between redshifts two and four, the study of the broad emission lines infer N/C ratios of up to 2, hence metallicities of up to 14 times solar \citep{hamann92}. Even for local quasars (z$<$0.1), the average metallicity derived from the N/C ratio is about twice solar \citep{hamann92}. Using the second method, \cite{arav07} measured absolute abundances for Mrk~279, which they found to be consistent with either solar or twice solar values. 

\subsubsection{Abundances from narrow absorption lines}

Narrow absorption lines were used to determine the abundances of the observed outflow by \cite{arav07} in the UV spectrum of Mrk~279. These authors measured the absolute abundances of C, N, and O to be C 2.2$\pm$0.7, N 3.5$\pm$1.1, and O 1.6$\pm$0.8, corresponding to 1.9$\pm$0.7, 3.6$\pm$1.1, and 1.3$\pm$0.8 on the \cite{lodders09} scale. If we compare our relative abundances to the absolute abundances measured for Mrk~279, we find that the C/O and N/O ratios are somewhat larger for Mrk~279 than for Mrk~509. However, their error bars are also larger, thus the abundance ratios measured for Mrk~279 \citep{arav07} are consistent within the error bars with both proto-solar values and the abundance ratios measured for Mrk~509. Using the VLT UVES spectrum of the AGN wind of the z=2.238 quasar QSO~J2233$-$606, \cite{gabel06} measured super-solar absolute abundances of C 4.0$-$5.0, N 12.6$-$15.9, and O 3.2$-$5.0. The abundance ratios thus range between 1 and 1.6 for C/O and 2.5 and 5 for N/O, indicating that there is a clear overabundance of N relative to O, not observed for Mrk~509.  

It would thus appear from the abundance measurements in Mrk~279 and Mrk~509 derived using high-resolution spectra, that in the local universe, the abundance ratios of the gas near the accretion disk are close to proto-solar/solar values. 

There have been several previous attempts to determine relative abundances using X-ray spectra, but generally the SNR of the data were too low for these measurements to be constrained. For earlier results, inaccuracies in the atomic data used also led to biased results, especially for Fe when the most recent di-electronic recombination rates were not used, as discussed by \cite{netzer04}. For instance, \cite{schmidt06} and \cite{nahar09} performed detailed atomic calculations of the di-electronic recombination rates. \cite{schmidt04} measured the wavelengths of the \ion{O}{v} lines, after \cite{steenbrugge03} had noted that the wavelength used for the 22~\AA~line was likely to be wrong. In their appendix 1, \cite{detmers11} list the atomic data that has been updated prior to the analysis of the RGS and LETGS spectra of Mrk~509. For the 40~ks RGS spectrum of NGC~3783, \cite{blustin02} measured the abundances to be consistent with solar, with the exception of Fe. Similar abundances were measured by \cite{steenbrugge03} for NGC~5548 using a 137~ks RGS spectrum, as well as by \cite{blustin07} for a 160~ks RGS spectrum of NGC~7469. \cite{holczer07} determined the abundances for IRAS~13349+2438 and NGC~3783 using the updated di-electronic recombination rates and determined abundances relative to Fe consistent with proto-solar ratios for O, Ne, Mg, Si, and S (NGC~3783 only). In Table~\ref{tab:wind}, we compare our relative abundances to those obtained by \cite{holczer07}, converting the abundances to the \cite{lodders09} abundance scale. Although some of these earlier abundance measurements suffer from large uncertainties, the determined abundances of AGN in the nearby universe using narrow X-ray absorption lines are broadly consistent with solar abundances.

\begin{table}[!h]
\begin{center}
\caption{Comparison of the relative abundances of Mrk~509 (see Table~\ref{tab:abun1}) with those derived for IRAS~13349+2438 and NGC~3783 \citep{holczer07}.}
\label{tab:wind}
\begin{tabular}{lccc}
\hline
\hline
       & Mrk 509       & IRAS~13349+2438   & NGC~3783 \\\hline
O/Fe   & 1.17$\pm$0.06 & 1.3$\pm$0.4       & 1.0$\pm$0.2 \\
Ne/Fe  & 1.31$\pm$0.17 & 1.3$^{+1.6}_{-0.4}$& 1.4$^{+0.3}_{-0.8}$\\
Mg/Fe  & 0.80$\pm$0.16 & 1.4$\pm$0.5       & 1.5$^{+0.4}_{-0.7}$\\
Si/Fe  & 1.53$\pm$0.60 & 2.2$\pm$0.75      & 1.8$^{+0.2}_{-0.6}$ \\
S/Fe   & 0.67$\pm$0.15 &                   & 1.0$^{+0.2}_{-0.5}$ \\\hline
\end{tabular}
\end{center}
\end{table}

\subsubsection{Abundances from broad emission lines}
We can also compare the abundances we derived, and those derived for the outflow in other nearby Seyfert~1 galaxies, with those derived for the broad emission-line region. Although the relationship between the absorbing gas and the BLR is unknown, since both are located near the accretion disk, their abundances should be similar, and we proceed under that assumption. In the approach that we adopt, some additional assumptions are necessary to determine the abundance, as in the optical and UV spectra only a handful of broad emission lines are observed. As a result, the ionisation structure of the gas is poorly constrained, and one thus has to rely on a model for the ionisation structure and to use lines with the same ionisation parameter, otherwise there is a degeneracy between the ionisation structure and abundance measured. In these studies, it is generally assumed that the N/O ratio is a proxy for O/H, which is itself a proxy for the metallicity Z, or that N/H is proportional to the metallicity squared \citep{hamann99}. The relation between N/O and metallicity was derived and is verified for \ion{H}{ii} regions \citep{searle71}. It was applied to quasar nuclei by \cite{shields76}, who also expanded the relationship by indicating that N/C should follow the same trend as N/O. \cite{shields76} and other early observers found a significant nitrogen overabundance relative to oxygen, using the \ion{N}{iv}] and \ion{N}{iii}] lines detected for a handful of nearby active galaxies; however, for the average spectrum of quasars they did measure N/O ratios of one-fourth to half solar. \cite{hamann93} showed that the \ion{N}{v}/\ion{C}{iv} and the \ion{N}{v}/\ion{He}{ii} ratios are less dependent on the ionisation parameter and are thus to be preferred. The \ion{N}{v}, \ion{C}{iv}, and \ion{He}{ii} lines are therefore currently the most widely used lines in determining the abundances from the broad emission lines.

Using the \ion{N}{v}/\ion{C}{iv} and \ion{N}{v}/\ion{He}{ii} ratios, high redshift quasars are found to have a range of abundances between a few to ten times solar \citep{hamann99}. These authors also derived a slight increase in abundance with redshift. \cite{hamann92,hamann93,hamann99} conclude therefore that quasar activity only becomes detectable after vigorous star formation, although the enrichment could be so rapid that star formation and QSO formation are coeval. 
This is in contrast to the earlier results, based on different emission lines and the study of lower redshift quasars, that generally found quasar abundances are sub-solar. Since both methods use gas that is formed at a very similar distance from the inner accretion disk that is assumed to have the same origin as the disk, the difference in abundance is due to either the assumed ionisation structure for the gas, because of galaxy evolution, or because one, or more, of the lines used in the abundance calculation is contaminated. Below we describe the observational evidence against such high metallicities at high redshifts, and argue that the \ion{N}{v} line used is most likely contaminated.

The high metallicities measured at high redshift imply that the initial mass function (IMF) was relatively flat during an intense period of star formation, which thus produced more high-mass stars than observed in the solar neighbourhood \citep{hamann99,hamann93}. This is inconsistent with the IMF derived for the eight most luminous and massive cluster galaxies in the nearby universe studied by \cite{vandokkum10}. These authors derive a steeper than Salpeter IMF, namely a power-law slope of $\sim$$-$3. 
This is much steeper than the power-law slope derived for the Milky Way disk, and implies that 80\% of stars in these galaxies are low-mass stars. \cite{vandokkum10} do point out that this IMF is applicable also to the central region, where the initial starburst should have occurred and that star formation caused by later mergers mostly affect the stellar mass function at larger radii. It thus seems that the high metallicities measured from the broad line region are inconsistent with the observed IMF in bright massive nearby cluster galaxies.

The possibility that the \ion{N}{v} line is contaminated was first suggested by \cite{turnshek88}, who found a possible correlation between the \ion{N}{v} broad emission line strength and the broad absorption lines measured in the AGN under study. \cite{krolik98} calculated the possible contribution of resonant scattering to the \ion{N}{v} line to be 0.4$-$0.5 times the Ly$\alpha$ flux. Finally, \cite{wang10} summarised the observational evidence which implies that indeed the \ion{N}{v} line is generally strongly contaminated by the resonance scattering of Ly$\alpha$ photons. Strong support is provided by the 10\% polarisation fraction of the \ion{N}{v} emission line \citep{lamy04}, while the finding that other broad emission lines are unpolarised indicates that at least part of the emission is scattered light. Furthermore, \cite{weymann91} found a difference between the strength of the \ion{N}{v} line in BAL and non-BALQSOs, with excess \ion{N}{v} emission extending to 10,000~km~s$^{-1}$, much broader than the broad emission lines. The contamination with resonant-scattered Ly$\alpha$ photons diminishes the reliability of the metallicities derived from the broad emission line region using the \ion{N}{v} line and might lead to spurious over-abundances. However, when using other line ratios the uncertainties caused by the difference in the line ratios with temperature, ionisation structure, density, and assumed spectral energy distribution ensure that the abundance determination is uncertain. Another complication of the abundance determination from broad emission lines is that the BLR is stratified, i.e. the ionisation parameter and hence line strengths change with distance from the black hole. Nevertheless, the abundances derived for the BLR are overwhelmingly using one zone models. Therefore, we conclude that the abundances determined for the BLR are unreliable, and especially when the \ion{N}{v} line is used.  


\subsection{Abundances in the Galactic centre}

Our own Galaxy is the only case where we can distinguish the abundances of the different components in the vicinity of the supermassive black hole. As a result, the abundances have been determined for both molecular clouds and \ion{H}{ii} regions and from both cool and hot stars. For the molecular clouds in the vicinity of the Galactic centre (GC), the derived iron abundance depends on the assumed process producing the observed X-ray emission, and ranges from 1.3$\pm$0.1 \citep{terrier10}, to 1.6$\pm$0.2 \citep{revnivtsev04}, and to 3.4 times the proto-solar value \citep{yusef-zadeh07}. Studying \ion{H}{ii} regions, \cite{giveon02b} determined a metallicity gradient for the Galaxy, which gives an abundance of the GC for Ne of 2.3$\pm$1.3 and for S of 0.68$\pm$0.27 times the proto-solar value. The Ne/S ratio of the \ion{H}{ii} regions near the GC is larger than that measured for Mrk~509, but both measurements are consistent. From the study of 11 giants and supergiants within 30~pc of the GC, \cite{cunha07} derived an average proto-solar iron abundance of 1.12$\pm$0.17 and an average O abundance of 1.9$\pm$1.1. Solar zero-age main-sequence abundances were derived from the study of five late-type Wolf-Rayet stars by \cite{najarro04}. Hence, most of the measurements of the absolute abundances in the immediate surroundings of Sgr~A* are consistent with proto-solar and twice proto-solar abundances. The O/Fe abundance ratio derived from cool giant and supergiant stars is consistent with the measured abundance ratio of Mrk~509.

\subsection{Metallicity of star-forming galaxies}
Here we compare the abundances measured for Mrk~509, with measurements of the metallicity in the interstellar medium (ISM) in nearby star-forming galaxies. We note that Mrk~279 is the only Seyfert 1 galaxy with reliable absolute abundances for the outflow. The gas near the accretion disk is due to inflow, hence the measured metallicity should be related to the metallicity of the ISM. In the comparison, we implicitly assume that the abundances of the gas falling toward the black hole do not differ significantly from the average abundances/metallicity of the host galaxy. Ideally, we would compare the measured nuclear abundances with the ISM abundances of the host galaxy of the Seyfert galaxies under study, but the metallicities of these galaxies are unavailable. Nor is there a known mass-metallicity relationship for either Seyfert or active galaxies. Therefore we use the mass-metallicity relationship derived by \cite{tremonti04} for star-forming galaxies, in addition assuming that the star-forming galaxies in the local universe have had a similar enrichment history to the host galaxies of Mrk~279 and Mrk~509. The metallicity is defined in \cite{tremonti04} as 12+log(O/H), where a value of 8.69 is the solar oxygen abundance determined by \cite{allendeprieto01}. To compare with the \cite{lodders09} abundances used, in the following we divide the values obtained using the \cite{tremonti04} relationship by 1.235. We do not have absolute abundances for Mrk~509, making a direct comparison impossible. However, for Mrk~279 the absolute O abundance \citep{arav07} is available and a comparison is possible. 

Mrk~509 is a very compact galaxy and therefore we assume that the host galaxy consists of only a bulge. We can calculate the mass of the galaxy to be $\sim$5$\times$10$^{10}$ M$_{\odot}$ using the relation given by \cite{magorrian98} and using the measured mass of the black hole M$_{BH}$ = 3$\times$10$^8$ M$_{\odot}$ \citep{mehdipour11}. The uncertainty in the mass of the black hole obtained from reverberation mapping is about 30\% \citep{peterson04}, and is higher for the mass of the whole galaxy owing to the spread in the Magorrian relation as well as our assumption that there is only a bulge in this clearly disturbed galaxy (see Fig. 8 of \citealt{kriss11}). However, the metallicity determination is only slightly affected by assuming masses of only 2$\times$10$^{10}$ or 8$\times$10$^{10}$~M$_{\odot}$.

For a galaxy mass of 5$\times$10$^{10}$ M$_{\odot}$, we obtain an O/H ratio that is 1.68 times the proto-solar value. For a mass of 2$\times$10$^{10}$~M$_{\odot}$ the O/H ratio is 1.65, and for a mass of 8$\times$10$^{10}$ M$_{\odot}$ it is 1.70 times the proto-solar value. The uncertainty in the assumed galaxy mass is rather large, but cannot explain the difference between the nuclear O abundance and the higher global abundance detemined for this galaxy. There is a spread in metallicity along the assumed relation of 0.1 dex, which is also unable to explain the super-proto-solar abundance. The fiber used in the \cite{tremonti04} study does not cover the whole galaxy, hence the integrated metallicity for the whole galaxy should be somewhat lower. The method of using nebular lines to determine metallicity might overpredict the metallicity by a factor of two \citep{kennicutt03}. If this were indeed the case then the metallicity of the host galaxy ISM would be about proto-solar. Determining absolute abundances and the relative abundances of different elements is difficult using nebular lines, for the same reasons as the measurements of abundances from broad emission lines is, which include the uncertainties in ionisation, temperature, and density of the gas, and that an unknown fraction of the heavy elements is locked up in dust. Since we only measure abundance ratios relative to O, the derived metallicity cannot be directly compared with our measured abundances for Mrk~509. However, we do find abundance ratios of the different elements are close to proto-solar values. 

\cite{arav07} did use the \cite{allendeprieto01} abundances, thus for Mrk~279 the metallicity could be directly compared to the absolute abundance. The mass of the host galaxy bulge of Mrk~279 is 2.29$\times$10$^{10}$~M$_{\odot}$ \citep{wandel02}, and the galaxy is classified as a S0 galaxy \citep{devaucouleurs91}. Assuming that the mass of the disk is 30\% of the bulge mass, the galaxy mass is somewhat lower than the bulge mass calculated for Mrk~509, $\sim$3$\times$10$^{10}$ M$_{\odot}$. Hence, the O/H ratio is 1.66, not correcting for a possible factor of two overprediction of the O/H ratio. \cite{arav07} measured a O/H abundance of 1.3$\pm$0.8, which is consistent with either proto-solar or twice the proto-solar metallicity. The large uncertainty in both the measured abundance and the galaxy mass determination, as well as the possible factor of two overprediction in metallicity, imply that the results are inconclusive, and clearly more reliable data are needed to improve this comparison.


\subsection{Abundances from the hot ISM}
An alternative way to study ISM abundances is to use X-ray emission lines from the hot ISM observed in some local elliptical galaxies. Owing to the larger number of ions and lines that can be observed, the problems in determining ionisation, temperature, and gas density are less severe than in the optical for the broad emission lines. However, these abundance determinations depend on the assumed thermal structure, as it is well known that a single temperature model predicts too low abundances if the gas has two temperatures \citep{trinchieri94,buote00}. Furthermore, in a few cases resonance scattering needs to be taken into account for the strongest iron lines. At the outskirts of the galactic X-ray halo, there might be contamination by either cluster or group emission, although, this does not affect the abundances derived for the centres of these haloes. \cite{ji09} studied ten local elliptical galaxies that have a bright X-ray halo to determine the abundances of the hot interstellar medium. The median relative abundances, for the elements that we also measure in Mrk~509, in the eight galaxies for which \cite{ji09} obtained an acceptable fit are (converting to \cite{lodders09} abundances) O/Fe = 0.66, Mg/Fe = 0.94, and Si/Fe = 1.6$\pm$0.6. These should be compared with our measurements of 1.17$\pm$0.06, 0.80$\pm$0.15, and 0.67$\pm$0.16, respectively, where the Si abundance is taken from the {\tt xabs} modelling of the LETGS spectrum. Although for the individual galaxies, \cite{ji09} determined a wide range of abundances there is a rather large difference in the O/Fe abundance compared to Mrk~509, for which the 75\% quartile is 0.77. The Mg/Fe and Si/Fe abundances are consistent with our results for Mrk~509. The oxygen abundance is generally observed to be low compared to the Fe and Mg abundances in the hot interstellar medium \citep{ji09,humphrey06}. The reason for this low O abundance is a bit of a mystery, but might indicate that massive stars lost more of their He-rich atmosphere before the supernova explosion (hereafter SN) occurred, and hence that there was less He burning in the outer envelope than current theories predict \citep{fulbright05}. For two galaxies of the ten studied by \cite{ji09}, the O/Fe ratio determined with the RGS in the inner 1$^{\prime}$ is close to the proto-solar ratio, hence our results are not necessarily inconsistent with the abundances measured for the hot interstellar medium of elliptical galaxies.

\subsection{Comparison with cluster abundances}
Finally we also compare the abundances to those for the intra-cluster medium (ICM) in clusters of galaxies. The large gravitational potential of the cluster allows it to retain the gas ejected from the galaxies forming the cluster. It has generally been assumed that from the intra-cluster abundances the relative ratio of SN type Ia versus core-collapse SN can be constrained; however, stellar winds do play a role \citep{simionescu09}. There has been some debate about the mixing of the differently enriched ejecta, but most metals generally appear to have a peak abundance in the core, even though the oxygen abundance is less centrally peaked than for instance the iron abundance \citep{werner06,simionescu09}. There is a spread in abundance ratios across the different clusters of galaxies studied, with for instance the Ne/Fe ratio varying between 0.73$\pm$0.18 in Hydra~A \citep{simionescu09} and 1.40$\pm$0.11 in 2A~0335+096 \citep{werner06}. The best fit Fe/O ratio varies between 0.63 to 2.5 in a sample of six clusters of galaxies (Hydra~A, Fornax, M~87, Centaurus, S\'{e}rsic~159-03 and A~1060), and depends on the distance from the cluster core at which the measurement was made. 

The gas that enriches the ISM, i.e. that from stellar winds and SN explosions, is also expected to enrich the ICM. Therefore, naively one would expect that the metal abundance ratios measured in the outflows of Seyfert~1 in galaxies and the ICM might be similar. The absolute abundances will differ because of the large reservoir of primordial gas in the ICM with which the enriched gas will mix. A possible difference is expected if the combined star-formation history and initial mass function (IMF) of the galaxies belonging to the central part of the cluster is different from the star-formation history and IMF of Seyfert galaxies. This would result in a different ratio of type Ia to core-collapse SNe, as well as alter the importance of stellar winds in the enrichment of the gas. The disturbed morphology of Mrk~509 likely implies that it has had a different star-formation history from that of the elliptical galaxies dominating in the centres of clusters. 

In Table~\ref{tab:cluster}, we compare the O/Fe, Ne/Fe, Mg/Fe, Si/Fe, S/Fe, and Ca/Fe ratios for Mrk~509 and the cores of the three clusters of galaxies Hydra~A \citep{simionescu09}, S\'{e}rsic~159-03 \citep{deplaa06}, and 2A~0335+096 \citep{werner06}. The data we use are for Hydra~A the results obtained from a fit to the inner 3$^{\prime}$, for S\'{e}rsic~159-03 a fit to the inner 4$^{\prime}$, and for 2A~0335+096 also the inner 4$^{\prime}$ for the O/Fe and Ne/Fe, but the inner 3$^{\prime}$ for Si/Fe, S/Fe, and Ca/Fe. The oxygen and neon abundances are taken from the RGS fit, in the case of Hydra~A the fit with {\it gdem}+1T, and gdem for the EPIC data are used \citep{simionescu09}; for S\'{e}rsic~159-03 and  2A~0335+096 the results, are from the {\it wdem} model \citep{deplaa06,werner06}. For Mrk~509, we use the RGS abundances measured using model 3 and assuming that all velocity components have the same abundances (see Table~\ref{tab:abun1}), with the exception of Si taken from the {\tt xabs} modelling of the LETGS spectrum. 
In clusters of galaxies, owing to the high temperature of the plasma and the extent of the source, broadening the emission lines detected by the RGS with the brightness profile of the source, the abundances are generally measured using the EPIC instruments. Hence, the relative abundances are measured with respect to iron, which is well constrained with the EPIC instruments due to the presence of the Fe-L and Fe-K complexes. The temperature is too high, even in cool-core clusters, to derive a reliable abundance for either nitrogen or carbon as these elements are fully ionised. The oxygen and neon abundances we list were determined from RGS spectra, as neon is blended with the Fe-L complex, and oxygen is blended with a local oxygen absorption in the EPIC data. In the comparison below we use the abundance ratios determined for the cluster core, as the oxygen abundance is generally poorly constrained at outer annuli, where there is no good RGS data available. In general the O/Fe ratio increases with distance from the core, but for the other elements the abundance ratios are consistent with no change with distance. However, the absolute abundance does decrease with distance from the cluster core. 

The O/Fe, Ne/Fe, and Si/Fe abundance ratios are larger in Mrk~509 than for the cores of clusters, while the Mg/Fe, S/Fe and Ca/Fe are within the range measured in the cluster cores. All abundance ratios are consistent with at least the abundance ratio measured in one of the three clusters Hydra~A, 2A~0335+096 and S\'{e}rsic~159-03. The higher oxygen, neon, and silicum abundances in the outflow of Mrk~509 reflects the greater importance to the enrichment of the ISM by both/either core-collapse SNe and/or stellar winds. The disturbed morphology of Mrk~509 implies that a higher core-collapse SN versus type Ia SN rate in the recent history of this galaxy is a likely explanation of the larger abundance ratios of O and Ne relative to Fe. Furthermore, stellar winds might be less likely to escape the potential of the galaxy, and therefore preferentially enrich the ISM, which would explain the overall higher abundance ratios observed in the outflow in Mrk~509 than in the cores of clusters of galaxies.  

\begin{table}[!h]
\begin{center}
\caption{Comparison of the relative abundances in Mrk~509 (see Table~\ref{tab:abun1}) with those derived for the cores of the cluster of galaxies Hydra~A, S\'{e}rsic~159-03, and 2A~0335+096 \citep{simionescu09,deplaa06,werner06}.}
\label{tab:cluster}
\begin{tabular}{lcccc}
\hline
\hline
       & Mrk 509       & Hydra A       & S\'{e}rsic~159-03 & 2A 0335+096   \\\hline
O/Fe   & 1.17$\pm$0.06 & 0.76$\pm$0.11 & 0.87$\pm$0.10     & 0.49$\pm$0.05 \\
Ne/Fe  & 1.31$\pm$0.17 & 0.84$\pm$0.20 & 0.73$\pm$0.12     & 0.85$\pm$0.08 \\
Mg/Fe  & 0.80$\pm$0.16 &               & 0.32$\pm$0.04     & 1.07$\pm$0.08 \\
Si/Fe  & 1.53$\pm$0.60 & 0.65$\pm$0.05 & 0.69$\pm$0.02     & 1.35$\pm$0.03 \\
S/Fe   & 0.67$\pm$0.15 & 0.58$\pm$0.05 & 0.52$\pm$0.02     & 1.14$\pm$0.03 \\
Ca/Fe  & 1.05$\pm$0.26 & 1.28$\pm$0.16 & 1.04$\pm$0.05     & 1.58$\pm$0.06 \\\hline
\end{tabular}
\end{center}
\end{table}   

\section{Conclusions}
We have studied the relative abundances determined from the X-ray spectrum of the outflow in Mrk~509, which has allowed an accurate and reliable determination of the abundance ratio for five elements C/O, N/O, Ne/O, Mg/O, S/O, and Fe/O abundances. For Si/O and Ca/O, less accurate abundances were obtained. We have found that the measured relative abundances of C, N, Ne, Mg, Si, Ca, and Fe relative to oxygen, are consistent with proto-solar ratios. Sulphur is slightly underabundant relative to the proto-solar ratios given by \cite{lodders09}. These abundances have been confirmed using absorption measure distribution modelling, albeit with larger errors. 

We compared our relative abundance determination to the absolute abundances measured for Mrk~279, another Seyfert~1 galaxy at a similar redshift. For Mrk~279, the absolute abundances were determined from the HST-STIS and FUSE spectra, and are consistent with the relative abundances measured for Mrk~509. The abundances derived for the warm absorber in IRAS~13349+2438 and NGC~3783 are consistent with our determined abundances and proto-solar ratios. We also compared our method and results with the abundance determination from the broad emission lines in quasars, and found the latter method to be unreliable because of the effect of the resonance-scattered Ly$\alpha$ photons at the wavelength of the \ion{N}{v} broad-emission line. We therefore compared our results with the abundances measured for the Galactic centre and found that within the error bars and uncertainties the results are consistent. In addition using the absolute abundances determined for Mrk~279 \citep{arav07}, we compared the abundances within the nucleus of these Seyfert galaxies with ISM metallicities predicted from the mass-metallicity relationship for star-forming galaxies \citep{tremonti04}. For Mrk~279, the absolute O abundance derived from UV absorption lines is consistent with the ISM metallicity, but for both measurements the error bars and uncertainties are large. As an alternative we compared our abundance ratios to those measured for the hot ISM in local elliptical galaxies. The Mg and Si abundance relative to Fe in Mrk~509 are consistent with those in the hot halo ISM, although we measure a larger O/Fe ratio than that of the hot ISM.

Finally we compared the abundance ratios for O/Fe, Ne/Fe, Si/Fe, S/Fe, and Ca/Fe to the ratios determined in the cores of three clusters of galaxies. We found that the outflow has a higher O/Fe and Ne/Fe abundance than the cluster cores, while the Si/Fe, S/Fe, and Ca/Fe are within the range of measured cluster abundance ratios. This indicates that core-collapse SNe and stellar winds have been more important to the enrichment of the interstellar medium of Mrk~509 than to the enrichment of the intra-cluster medium of these clusters. Likely explanations are that stellar winds have not escaped the potential of the galaxy as readily as SNe ejecta, and that owing to a disturbance of the host galaxy of Mrk~509 the recent star-formation history is likely different from that of cD galaxies residing at the cores of clusters.

\begin{acknowledgements}
This work is based on observations obtained with XMM-{\it Newton}, an ESA science mission with instruments and contributions directly funded by ESA Members States and the USA (NASA). KCS would like to thank ESO Chile and SRON for their hospitality and acknowledges support from Comit\'{e} Mixto ESO - Gobierno de Chile. SRON is supported financially by NWO, the Netherlands Organization for Scientific Research. The Technion team was supported by a grant from the ISF. GAK gratefully acknowledges support from NASA/XMM-Newton Guest Investigator grant NNX09AR01G. MM acknowledges the support of a PhD studentship awarded by the UK Science \& Technology Facilities Council (STFC). NA gratefully acknowledges support from NASA/Chandra Guest Investigator grant GO0-11113X, and from NASA/HST Guest Investigator grant HST-GO-12022. POP acknowledges financial support from the french spatial agency CNES and from the french GDR PCHE. GP acknowledges support via an EU Marie Curie Intra-European Fellowship under contract no. FP7PEOPLE-2009-IEF-254279. GP and SB acknowledge financial support from contract ASI-INAF n. I/088/06/0.

\end{acknowledgements}  

\bibliography{references_abs}

\end{document}